\documentclass[11pt,twoside]{article}  
\usepackage{asp2006}
\usepackage{adassconf}
\usepackage{hyperref}

\begin{document}   

%

\paperID{O14.3}

%

\title{Status of GDL - GNU Data Language}
       
%
%
%
%
%

\markboth{Coulais et al.}{Status of GDL - GNU Data Language}

%

\author{A. Coulais}
\affil{LERMA, Obs. de Paris, ENS, UPMC, UCP, CNRS, Paris, France}

\author{M. Schellens\altaffilmark{1}}
\author{J. Gales}
\affil{Goddard Space Flight Center, Greenbelt, MD, USA}

\author{S. Arabas}
\affil{Institute of Geophysics, University of Warsaw, Poland}

\author{M. Boquien}
\affil{University of Massachusetts,
Dep. of Astronomy, Amherst, MA, USA}

\author{P. Chanial}

\author{P. Messmer, D. Fillmore}
\affil{Tech-X GmbH, Zurich, Switzerland; Tech-X Corp, Boulder, CO, USA}

\author{O. Poplawski}
\affil{Colorado Div. (CoRA) of NorthWest Res. Ass. Inc., Boulder, CO, USA}

\author{S. Maret}
\affil{LAOG, Obs. de Grenoble, UJF, CNRS, Grenoble, France}

\author{G. Marchal\altaffilmark{2}, N. Galmiche\altaffilmark{2}, T. Mermet\altaffilmark{2}}


\altaffiltext{1}{Head of the project} 
\altaffiltext{2}{Former students at LERMA CNRS and Observatoire de Paris}


\contact{Alain Coulais}
\email{alain.coulais@obspm.fr}

%
%

\paindex{Coulais, A.}
\aindex{Schellens, M.}
\aindex{Gales, J.}
\aindex{Arabas, S.}
\aindex{Boquien, M.}
\aindex{Chanial, P.}
\aindex{Messmer, P.}
\aindex{Fillmore, D.}
\aindex{Poplawski, O.}
\aindex{Maret, S.}
\aindex{Marchal, G.}
\aindex{Galmiche, N.}
\aindex{Mermet, T.}


\keywords{software!languages,
data processing!reduction,
methods!numerical,
services!data processing,
visualization}


\setcounter{footnote}{2}
\enlargethispage{0.1\baselineskip}

\begin{abstract}
Gnu Data Language (GDL) is an open-source interpreted language aimed
at numerical data analysis and visualisation. It is a free
implementation of the Interactive Data Language (IDL) widely used in
Astronomy. GDL has a full syntax compatibility with IDL, and includes
a large set of library routines targeting advanced matrix
manipulation, plotting, time-series and image analysis, mapping, and
data input/output including numerous scientific data formats. We will
present the current status of the project, the key accomplishments,
and the weaknesses - areas where contributions are welcome!
\end{abstract}


\section{Dependencies}

GDL is written in C++ and can be compiled on systems with GCC ($\ge$ 3.4)
and X11 or equivalents. The code, under
\htmladdnormallink{GNU GPL}{http://www.gnu.org/licenses/gpl.html}, 
is hosted by
\htmladdnormallink{SourceForge}{http://sf.net/projects/gnudatalanguage/}.
The library routines make use of numerous open-source libraries including:
\htmladdnormallink{readline}{http://tiswww.case.edu/php/chet/readline/rltop.html},
the GNU Scientific Library
(\htmladdnormallink{GSL}{http://www.gnu.org/software/gsl/}), the
\htmladdnormallink{PLplot plotting
  library}{http://plplot.sourceforge.net/},
a Fourier transform package
(\htmladdnormallink{FFTw}{http://www.fftw.org/}), and others.
Since recently (GDL 0.9rc4 release) GDL features multi-threaded matrix 
operations if compiled using an OpenMP-enabled compiler (e.g. GCC $\ge$ 4.2.)

PLplot and GSL are the only mandatory dependencies.  

Data input/output is managed using
\htmladdnormallink{ImageMagick}{http://www.imagemagick.org/},
\htmladdnormallink{NetCDF}{http://www.unidata.ucar.edu/software/netcdf/},
\htmladdnormallink{HDF}{http://hdf4.org/products/hdf4/} and 
\htmladdnormallink{HDF5}{http://hdf4.org/HDF5/} libraries.
FITS files can be read and written
using the \htmladdnormallink{Astron Library}{http://idlastro.gsfc.nasa.gov/}.
GDL features a Python
bridge (Python code can be called from GDL, GDL can be compiled as a
Python module).

\section{Platforms, packages and compilation}

GDL runs on most recent Linux and *BSD systems, and also on Mac OSX and
OpenSolaris. x86 and x86\_64 are the key supported architectures, 
successful compilations on other architectures have been reported.

Pre-compiled packaged versions of GDL are available for several
operating systems including Mac OSX (e.g. via Macports), Debian and
\htmladdnormallink{Ubuntu}{http://doc.ubuntu-fr.org/gdl},
Fedora and Red Hat, Gentoo, ArchLinux and FreeBSD.

The source code compiles smoothly on most Linux distributions, *BSD,
recent OSX\footnote{On Mac OSX, which is an
  OS GDL does have significant number of requests from users, success have been
  reported on 10.4, 10.5 and 10.6. A tutorial for compilation on OSX
  which try to include the last tricks can be found here:
  \makeURL{http://aramis.obspm.fr/\~coulais/IDL\_et\_GDL/GDLonOSX\_10.5.6.html}}
versions and some other UNIX systems (e.g. OpenSolaris),
as long as some caveats are avoided. 

Since the pre-compiled packages happen to be out of date and don't
include the latest bug-corrections and improvements, we strongly advice
to try compiling GDL from source. The source code can be obtained from 
\htmladdnormallink{SourceForge}{http://sf.net/projects/gnudatalanguage/}
where GDL development is hosted. 
The most recent additions to GDL are readily available at the 
\htmladdnormallink{CVS repository}{http://gnudatalanguage.cvs.sourceforge.net/viewvc/gnudatalanguage/gdl/}
\footnote{see \makeURL{http://sf.net/projects/gnudatalanguage/develop} for details}.

\section{Useful libraries}

Large parts of the \htmladdnormallink{Astron
  Library}{http://idlastro.gsfc.nasa.gov/} are working well in GDL,
including the FITS part.
The \htmladdnormallink{MPFIT}{http://www.physics.wisc.edu/\~craigm/idl/fitting.html}
library (Robust non-linear least squares curve fitting based on MINPACK-1) does work with GDL.
IDL save files can be read and written using the external 
\htmladdnormallink{CMSVLIB}{http://www.physics.wisc.edu/\~craigm/idl/cmsave.html} library.
Success have been reported with the Wavelet Library.

Due to limited number of graphical keywords currently available,
and also due to the limited achievements of the Postscript output,
some famous public libraries for complex graphical outputs (X11
and PS) are clearly limited in functionality.

\section{Contributions}

Contributions are very welcome. A large number of bug-reports and improvements
are coming from anonymous users, posting directly to the contributors
or posting on the SourceForge forum.  We really appreciate code
donation under GNU GPL.  There is a strong demand for more packaging effort.
Development is clearly driven by end users (you), except for very
difficult tasks (MEDIAN code development and testing was paid, based
on T. Huang et al. (1979)). Please do not hesitate to report bugs, regressions,
compilation problems, feature requests (e.g. missing keywords), or other
comments -- you can do it also anonymously at GDL SourceForge page 
(bug tracker or forum).  Please do help in testing GDL by
using the CVS version. Finally, please do not be impatient: GDL is developed
by a team of volunteers -- often busy with other tasks. 

\section{Weakness}

GDL suffers from several weakness, the main one maybe is that
none of the main contributors are full time on the project.

\begin{itemize}
\item  \textbf{Documentation}
Because of (1) the huge amount of on-line documentation concerning IDL,
(2) the goal to be as close as possible to IDL features,
(3) we use many external features (e.g. mathematical functions in GSL )
we lack a clear pure GDL documentation.
Nevertheless, there are some useful documents available on the web,
covering different aspects of the project, from
installations problems to usage in dedicated field.

\item  \textbf{Widgets}
Widgets are not available in GDL now.
An implementation based on wxWidgets library is being developed
but it is still in very early stage of development.

\item  \textbf{Graphical outputs}
The main weakness now is the Postscript output.
The current plotting features do not support publication-quality
figures. Contributions on this part will be very welcome.
Performance of simple plots (PLOT, OPLOT, PLOTS) in X11 using PLplot
are comparable to IDL ones. SURFACE and TV are not as fast,
especially trough network. For CONTOUR and SURFACE, PLplot
does not provide exact equivalent to IDL outputs.

\item  \textbf{Obsolete packages}
Another clear problem comes from the gap between the CVS version
and the versions available in main Linux distributions. The closest
version now is the Fedora one, which usually include
important corrections in the CVS.

\end{itemize}

\section{Usage}

As for many free software packages, we don't know if GDL is widely
used or not, by who and in which field (Astronomy, Geophysics,
Medicine ...)  but we have some indicators that this project does have
users !  We received messages worldwide, mostly from people working in
Astronomy. The greater the number of \htmladdnormallink{available
  features}{http://aramis.obspm.fr/\~coulais/IDL\_et\_GDL/Matrice\_IDLvsGDL\_intrinsic.html}
and the simpler the compilation procedure, the less the number of
messages.

The known examples of usage include: a master course at the
Paris Observatory ($\sim$20 students per year), a master course at the
University of Warsaw, refereed papers where computations were done
with GDL (e.g. : Koleva \& al 2009), computations for MSc and PhD
theses.

There are some documented examples of experimental extensions to GDL
written for specific research purposes. Jaffey et al. (2008) described
a web-based interactive data analysis tool based on GDL, while 
Jaffey et al. (2009) presented application of a GPU-accelerated version 
of GDL in solar-physics calculations.

\section{Conclusion}

The core components of GDL (i.e. interpreter, library routines API,
key data manipulation and plotting functionality) are stable and do
not pose efficiency problems (no significant discrepancy from IDL
performance). 
Large number of routines are available,
several widely used external libraries (Astron, MPfit)
can be used in GDL.

We hope to consolidate the users community, to gather feedback in form
of bug reports, feature requests, test routines, documentation and
patches (several GDL modules have been provided by scientists who
wrote the functions for their own work).

The current status is stable and complete enough for numerous applications
but still much work is needed.
The major axis on development for the next year are:
\begin{itemize} 
\item aggregating a more efficient community
\item giving pre-compiled versions in major Linux distributions closer
  to the CVS version
\item finishing a whole test suite (like the GSL one) to avoid
regression and bugs
\item having a wider set of graphical keywords
\item better Postscript output (or other format(s) for graphical outputs)
for publications
\item development of documentation
\end{itemize}

\acknowledgments

We are grateful to many anonymous contributors who spend time and
energy on testing and improving GDL. We appreciate all the feedback we
got so far from GDL users.
Alain C. would like to thanks his former students and also 
LERMA for financial support.

\end{document}